\documentclass[aps,prb,showpacs,groupedaddress]{revtex4}
\usepackage{amsmath,graphicx}
\usepackage{epsfig}
\usepackage{bm}
\usepackage[next]{inputenc}

\begin{document}
\title{Effect of anisotropy on the field induced quantum critical properties of the three dimensional s=1/2 Heisenberg model}
\author{H. Rezania}
\email[]{hamedrzn8@gmail.com}
\affiliation{Physics Department, Razi Univerity, Kermanshah, Iran}
\author{A. Langari}
\affiliation{Physics Department, Sharif University of Technology, Tehran 11155-9161, Iran}
\begin{abstract}
The field induced quantum critical properties of the three dimensional spin-1/2 anisotropic antiferromagnetic Heisenberg model has been studied.
We have investigated the quantum phase transition between the spiral order and field induced ferromagnetic order by means of Bose-Einstein condensation of magnons in terms of a bosonic representation. 
The effect of in-plane anisotropy on the critical properties has been studied
via the bosonic model by Green's function approach. We have found an analytic expression
for the gap exponent in addition to numerical results for the critical magnetic field in terms
of anisotropy parameter.
The in-plane anisotropy breaks the U(1)  symmetry explicitly which changes
the universal behavior by a drastic change on the gap exponent. Moreover, the critical
magnetic field depends strongly on the in-plane anisotropies. The divergence of the 
transverse structure factor
at the antiferromagnetic wave vector confirms the onset of the
magnetic order which scales with the negative value of gap exponent as the magnetic field
approaches the critical one.
The transverse staggered magnetization as an order parameter vanishes
with exponent $\beta=0.5$ when the magnetic field reaches its critical value in low field region.

\end{abstract}
\date{\today}

\maketitle
\section{introduction}
Quantum phase transition \cite{sachdev,vojta} is an interesting topic for both theoretical and
experimental condensed matter research activities. 
This phase transition is found at the zero temperature based on variation of non thermal control parameter such as magnetic field or hole doping. Quantum phase transition (QPT) occurs
at the quantum critical point where quantum fluctuations destroys a long range order of 
the model at absolute zero temperature. 
One of the novel species of quantum phase transition is field-induced-magnetic phase transition that 
can be occurred in insulating antiferromagnetic systems \cite{gia} like transition metal oxides and local spin systems. 
This kind of QPT has been observed in the copper halide Cs$_{2}$CuCl$_{4}$ which is an insulator
and each Cu$^{2+}$ carries a spin of $1/2$.

Cs$_{2}$CuCl$_{4}$ can be described as a quasi-two-dimensional spin 1/2 antiferromagnet on a triangular lattices
({\it bc} plane) weakly coupled along the crystallographic {\it a} direction.\cite{coldea}
The crystal field effects quench the orbital angular momentum; however, the anisotropic effect is still significant on the phase transition. 
The layered crystal structure confines the main superexchange routes to neighboring spins in the {\it bc} plane. 
According to the above facts the magnetic properties of this material can be described by the
antiferromagnetic Heisenberg model.
The magnetic field is  applied perpendicular to the {\it bc} plane which 
adds a Zeeman term to the model.
For magnetic fields ($B$) close to the critical field ($B_{c}$), Zeeman term competes with the  spin exchange 
interaction and system enters a field induced ferromagnetic state \cite{viktor1,nikuni}.
The field induced ferromagnetic phase ($B>B_c$) has gapped quasi-particles, gapped magnons. 
The field induced gap vanishes at $B_{c}$ when the magnetic field is reduced and 
the magnetic ordering for transverse component of spins sets up.
This latter state (for $B<B_c$) is named spiral long range order. 
In this work we have studied the mentioned quantum phase transition based 
on Bose-Einstein condensation of magnons via a bosonic gas model\cite{matsubara}. 
Bloch\cite{bloch} applied the Bose-Einstein quantum statistics to the excitations in solids
which gives the basic notion to relate Bose-Einstein of magnons to the magnetic ordering 
in the original spin model\cite{nikuni,viktor2}.

The isotropic Heisenberg model with longitudinal applied magnetic field 
has been studied by
theoretical and numerical methods.
The isotropic model on cubic lattice has been investigated by numerical quantum Monte-Carlo method 
at finite temperature which gives the phase boundary between spiral order and induced ferromagnetic state\cite{nohadani}. A theoretical approach based on Bose condensation of magnons 
for the isotropic model on triangular lattice has been studied in Ref.[\onlinecite{viktor1}]. 
Furtheremore, the experimental data
for specific heat indicate that the $\lambda$ like anomaly peak appears in the behavior of specific heat versus temperature for $B<B_{c}(T=0)$\cite{viktor2}.

In the general case, the spin model Hamiltonian can include spatial anisotropies
in the exchange coupling between nearest neighbor spins. This property is related to the 
existence of easy axes magnetization due to
crystalline electric field and spin-orbit coupling.
Dzyaloshinskii-Moriya (DM) interaction with a DM vector in a specific direction 
establishes easy-plane spin anisotropy in the Cs$_{2}$CUCl$_{4}$\cite{moriya}. 
Anisotropy due to DM interaction violates SU(2) symmetry of the isotropic Hamiltonian although U(1) symmetry corresponding to spin rotation
around DM vector is still present. However, the spin-orbit coupling may induce the anisotropy
in the {\it bc} easy-plane which reduces the U(1) to Z(2) symmetry.
Therefore, Goldstone theorem \cite{auerbach} can not be applied in such cases, because it is 
applied for Hamiltonian with a continuous symmetry.
In other words, the excitations of this model are not Goldstone modes.

In this paper, we intend to find the effect of in plane anisotropy on the 
critical point and the transverse spin structure factor close to the field induced QPT.
We have considered the fully anisotropic spin 1/2 Heisenberg model in the presence of 
a longitudinal field on a cubic lattice. We anticipate that the general behavior 
for cubic lattice is also valid for the triangular one. Moreover, the study on cubic lattice 
reduces the complexity of calculations which will be the route to investigate the triangular case.
In addition the results on cubic lattice can be applied to the field-induced magnetic
phase transition of TlCuCl$_3$. 
We have implemented the hard core boson transformation for spin operators which gives the 
excitation spectrum in terms of  many body calculations for bosonic gas \cite{gorkov}. 
We have used Brueckner approach\cite{fetter}
to find the bosonic self energy to get the magnon dispersion relation. 
The quantum critical point is approached where the magnon spectrum becomes gapless. 
We have found an analytic expression for the gap exponent in terms of the anisotropy parameter.
Our results show that a small amount of in-plane anisotropy changes the gap exponent drastically
which is the witness for the change in universal behavior. We have also found the dependence
of critical magnetic field on the anisotropy parameter which is also justified by the
divergence of the transverse structure factor at the antiferromagnetic wave vector.
The divergence of the  in-plane magnetic susceptibility obeys an algebraic power law
with an exponent equals to the negative of gap exponent as the magnetic field approaches
the critical one. Moreover, the vanishing of the staggered magnetization is given 
be the exponent $\beta=0.5$ in the mean field approximation.

\section{Anisotropic spin Hamiltonian}
The most general effective Hamiltonian to describe magnetic insulating matter due to exchange interaction between the spin of localized electrons can be written by
\begin{eqnarray}
\mathcal H=\frac{1}{2}\sum_{\langle ij\rangle}\sum_{\alpha=x,y,z}J_{ij}^{\alpha}S^{\alpha}_{i}S^{\alpha}_{j}-g\mu_{B}B\sum_{i}S_{i}^{z},
\label{e1}
\end{eqnarray}
where $g\simeq 2.2$, $\mu_B$ is the Bohr magneton and $B$ is the magnetic field.
The localized spins are located on the cubic lattice structure with nearest neighbor exchange interaction.
The effect of spin orbit coupling
is generally entered to the fully anisotropic exchange couplings $J_{ij}^{\alpha}$.
The exchange anisotropy is defined by parameter $\nu$ with the following relations
\begin{eqnarray}
J^{x}=J(1+\nu)\;\;,\;\;J^{y}=J(1-\nu)\;\;,\;\;J^z=J,
\label{e2}
\end{eqnarray}
where the scale of energy $J$ is set to one. 
The above type of exchange interaction implies an anisotropy in both in plane and axial directions
where the symmetry has be reduced to Z(2).

\section{Hard core representation of spin Hamiltonian and bosonic Green's functions}

As mentioned in the introduction we intend to describe the field induced QPT in terms of 
Bose-Einstein condensation of magnons. Our approach is similar to what we have implemented
in Ref.\onlinecite{rezania2008} to study the quantum critical properties of the Kondo-necklace
model.
In the first step the spin Hamiltonian is
mapped to a bosonic model.
This is done by the hard core boson representation  given by: $S^{+}_{i}\longrightarrow a_{i},\;S^{-}_{i}\longrightarrow a^{\dag}_{i}$ and $S^{z}_{i}=1/2-a^{\dag}_{i}a_{i}$,
where $a_i$ and $a_i^{\dagger}$ are boson annihilation and creation operators, respectively.
The SU(2) algebra of spin operators is retrived from bosonic algebra of $a_i$ and $a_i^{\dagger}$ operators with the  hard core constraint, i.e only one boson can occupy a single site of lattice. 
The constraint is added to the Hamiltonian by an on-site infinite  repulsion among bosonic particles. The
resulting Hamiltonian in terms of bilinear ($\mathcal H_{bil}$) and interacting ($\mathcal H_{int}$) 
parts is given by
\begin{eqnarray}
\mathcal H_{bil}=\frac{1}{2}\sum_{i,j}\{J [a_{i}^{\dag} a_{j} +\frac{1}{2}\nu 
(a_{i}^{\dag} a_{j}^{\dag}+a_{i} a_{j})]-J^{z} a_{i}^{\dag} a_{i}\}
+g\mu_{B}B\sum_{i}a_{i}^{\dag}a_{i}, 
\label{a3}
\end{eqnarray}
\begin{eqnarray}
\mathcal H_{int}=\mathcal U\sum_{i}a_{i}^{\dag}a_{i}^{\dag}a_{i}a_{i} +
\frac{1}{2}\sum_{ij}J^z a_{i}^{\dag}a_{i}a_{j}^{\dag}a_{j}.
\label{a4}
\end{eqnarray}
The bilinear Hamiltonian in the Fourier space representation is
\begin{eqnarray}
\mathcal{H}_{bil}=\sum_{\bf k}\{A_{\bf k}a_{\bf k}^{+} a_{\bf k}+
\frac{B_{\bf k}}{2}(a_{\bf k}^{+} a_{\bf -k}^{+} + a_{\bf -k}a_{\bf k} )\}, \label{e6}
\end{eqnarray}
\begin{eqnarray}
A_{\bf k}&=&[J_{\bf k} - J_0^z +g\mu _{B}B],\nonumber\\
B_{\bf k}&=&\nu J_{\bf k}.
\end{eqnarray}
It is defined $J_{\bf k}=J \sum_{\alpha=x,y,z} \cos(k_{\alpha})$ and $J_0^z=3J$.
The wave vectors $k_{\alpha}$ are considered in the first Brillouin zone. 
The effect of hard core repulsion part ($\mathcal U \rightarrow \infty$) 
of the interacting Hamiltonian in Eq.(\ref{a4}) is dominant 
compared with the second quartic term. Thus, it is sufficient to take into account the effect of 
hard core repulsion on the magnon spectrum and neglect the second quartic term. 
The interacting part of Hamiltonian in terms of Fourier transformation of bosonic operators is given by
\begin{eqnarray}
 \mathcal{H}_{int}=\mathcal{U}\sum_{k,k',q}a^{\dag}_{
k+q}a^{\dag}_{k'-q }a_{k'}a_{k} \;.
\label{e4}
\end{eqnarray}
The bilinear Hamiltonian  is simply diagonalized by the unitary Bogoliuobov transformation  
to the new bosonic quasi particle operators $\alpha_{\bf k}$ and $\alpha_{\bf k}^{\dag}$, 
[\onlinecite{rezania2008}],
which is given by
\begin{eqnarray}
\mathcal {H}_{bil}&=&\sum_{\bf k}\omega_{\bf k}\left( \alpha_{\bf k}^{\dag}\alpha_{\bf k} +1/2\right),\nonumber\\
\omega^{2}_{\bf k}&=&A_{\bf k}^2-B_{\bf k}^2,
\end{eqnarray}
and the Bogoliubov coefficients are
\begin{eqnarray}
 u^{2}_{\bf k} (v^{2}_{\bf k})=(-)\frac{1}{2}+\frac{A_{\bf k}}{2\omega_{\bf k}}.
\label{e64}
\end{eqnarray}
Although the bilinear part of Hamiltonian is diagonal in the new bosonic 
($\alpha_{\bf k}, \alpha_{\bf k}^{\dag}$) representation, to avoid the complexity of 
the calculations for the hard core repulsion term the Green's functions will be 
calculated in the original boson operators ($a_k, a_k^{\dagger}$).
In the original boson representation, $\mathcal {H}_{bil}$ includes the pairing term between magnons
which requires 
both anomalous and normal Green's functions to be considered.
More explanations of the detailed calculations can be found in 
Ref.\onlinecite{rezania2008}.
Finally, the self-energy is expanded in the low energy limit which gives the single particle part of Green's function ($G_{n}^{sp}$),
\begin{equation}
G_{n}^{sp}(k,\omega)=
\frac{Z_{k}U_{k}^{2}}{\omega-\Omega_{k}+i\eta}-\frac{Z_{k}V_{k}^{2}}{\omega+\Omega_{k}-i\eta},
\label{e330}
\end{equation}
where the renormalized triplet spectrum ($\Omega_{k}$), the renormalized single particle 
weight constants ($Z_{k}$) and renormalized Bogoliuobov coefficients ($U_{k}, V_{k}$) are given by
\begin{eqnarray}
\Omega_{k}&=&Z_{k}\sqrt{[A_{k}+\Sigma_{n}(k,0))]^{2}-[B_{k}+\Sigma_{a}(k,0)]^{2})},
\nonumber\\
&&Z_{k}^{-1}=1-(\frac{\partial \Sigma_{n}}{\partial \omega})_{\omega=0},\nonumber\\
&&U_{k}^{2} (V_{k}^{2})=(-)\frac{1}{2}+\frac{Z_{k}[A_{k}+\Sigma_{n}(k,0)]}{2\Omega_{k}}.
\label{e340}
\end{eqnarray}
The renormalized weight constant is the residue of the single particle pole in the
Green's function. In the next step we will take into account the effect of hard core
repulsion on the magnon spectrum.

\section{Effect of hard core repulsion on the magnon spectrum}
The density of the magnons is obtained from the normal Green's functions
\begin{eqnarray}
n_{i}=\langle a^{\dag}_{i}a_{
i}\rangle
=\frac{1}{N}\sum_{k}v^{2}_{k},
\label{e131}
\end{eqnarray}
where $N$ is the number of the spins in the cubic lattice. 
In the vicinity of the critical field ($B^{0}_{c}$) and at the zero temperature the density of excited magnons is negligible\cite{nikuni}. 
Since the Hamiltonian $\mathcal{H}_{int}$ in Eq.(\ref{e4}) is short ranged and $\mathcal{U}$ is large, 
the Brueckner approach (ladder diagram summation)\cite{gorkov,fetter} can be applied
for the low density limit of magnons.
The interacting normal Green's function is obtained by imposing the hard core boson 
repulsion, $\mathcal{U}\rightarrow \infty$. 
Firstly, the scattering amplitude (t-matrix) $\Gamma(k_{1},k_{2};k_{3},k_{4})$ 
of magnons is introduced where $k_{i}\equiv(\textbf{k},(k_{0}))_{i}$. 
The basic approximation made in the derivation of $\Gamma(K)$ is
that we neglect all anomalous scattering vertices, which are
presented in the theory due to the existence of anomalous Green's functions.
According to the Feynman rules\cite{fetter} in momentum space at zero temperature, 
the scattering amplitude is calculated
(see Fig.1 of Ref.\onlinecite{rezania2008}).
By replacing the  noninteracting normal Green's function 
in the Bethe-Salpeter equation 
and taking the 
limit $\mathcal{U}\longrightarrow\infty$ we obtain 
the scattering matrix in the form 
\begin{eqnarray}
\Gamma(\textbf{K},\omega)=-\Big(\frac{1}
{(2\pi)^{3}}
\int d^{3}Q\frac{u^{2}_{\textbf{Q}}u^{2}_{\textbf{K}-\textbf{Q}}}
{\omega-\omega_{\textbf{Q}}-\omega_{\textbf{K}-\textbf{Q}}}
-\frac{v^{2}_{\textbf{Q}}v^{2}_{\textbf{K}-\textbf{Q}}}
{\omega+\omega_{\textbf{Q}}+\omega_{\textbf{K}-\textbf{Q}}}
\Big)^{-1}.
 \label{e178}
\end{eqnarray}
According to Fig.2 of Ref.\onlinecite{rezania2008} and after 
some calculations the normal self-energy is obtained 
in the following form
\begin{eqnarray}
\Sigma^{\mathcal{U}}_n(\textbf{k},\omega)&=&\frac{2}{N}\sum_{p} v_{\textbf{p}}^{2}\Gamma(\textbf{p}+\textbf{k},\omega-\omega_{\textbf{p}}).
 \label{e211}
\end{eqnarray}
 In the dilute gas approximation there are other diagrams which are formally at most linear in 
the density of magnons. However, the magnon densities are very small and the contributions of 
such terms are numerically smaller than Eq.~(\ref{e211}).
 We should also consider the anomalous self-energy related to $H_{\mathcal{U}}$ which 
exists in the vertex function.
The anomalous self-energy has a vanishing contribution.


\section{the gap exponent}
Close to the quantum critical point, the excitation gap ($\Delta$) in the field-induced ferromagnetic phase vanishes according to the following power law behavior
\begin{eqnarray}
\Delta\sim |B-B_{c}|^{\phi},
\label{gapscaling}
\end{eqnarray}
where $B_{c}$ is the critical magnetic field and $\phi$ is the gap exponent which is related to universality class of the quantum critical point. 
The quantum critical point corresponds to the vanishing of magnon spectrum at the
antiferromagnetic wave vector $Q_{AF}=(\pi, \pi, \pi)$. The magnon spectrum  close to 
the antiferromagnetic wave vector ($Q_{AF}$) is approximated by
\begin{eqnarray}
\omega_{k}=\sqrt{\Delta^{2}+c^{2}(k-Q_{AF})^{2}},
\end{eqnarray}
where $c$ is the spin wave velocity. 
The spin wave velocity ($c$) is obtained numerically from the excitation spectrum.

In the first step, we calculate the variation of the self-energy related to 
$H_{\mathcal U}$ which is given by
\begin{eqnarray}
\delta \Sigma^{\mathcal U}(Q_{AF})
&=&\frac{2}{N}\sum_{k}\delta v_{k}^{2}\Gamma(k+Q_{AF},-\omega_{k})+\frac{2}{N}\sum_{k}v_{k}^{2}\delta \Gamma(k+Q_{AF},-\omega_{k}).
\label{e750}
\end{eqnarray}
The main contribution to the first integral in  Eq.~(\ref{e750}) comes from the
small momenta $q\sim \Delta/c\ll1$ where $q\equiv k-Q_{AF}$ since
\begin{eqnarray}
\delta v_{k}^{2}=\frac{1}{2} \Big(\frac{\delta A_{k}}{\omega_{q}}+A_{k}\delta [\frac{1}{\omega_{k}}]\Big)\approx -\frac{A_{Q_{AF}}^{c}\Delta^{2}}{2(\Delta^{2}+c^{2}q^{2})^{3/2}}.
\label{e800}
\end{eqnarray}
Taking into account the first correction to the magnon density, the vertex function can be written for small $q$ (see Ref.[\onlinecite{shevchenko}])
\begin{eqnarray}
 \Gamma(q,-\omega_{q})\approx\Gamma^{c}_0 [1+\frac{{\Gamma^{c}_0}A_{k=0}^{c}}{4\pi c^{2}}\ln q],
\label{e900}
\end{eqnarray}
where $\Gamma^{c}_0\equiv\Gamma^c(k=0)$ and the value of quantity $X$ at the critical point is defined by $X^c$.
The substitution of Eq.(\ref{e900}) in  Eq.(\ref{e750}) and replacing
$q\simeq \Delta/J$ in the first integral of Eq.(\ref{e750}),
we find that
\begin{eqnarray}
 \delta \Sigma^{\mathcal U}(\pi,\pi)=-\frac{A_{Q_{AF}}^{c}\Delta^{2}}{8\pi^{2} c^{3}}\Gamma^{c}_0[1+\frac{{\Gamma^{c}}_0A_{k=0}^{c}}{4\pi c^{2}}\ln\frac{\Delta}{J}]+\Gamma^{\prime}n_{b}\delta B,
\label{e950}
\end{eqnarray}
where $\Gamma^{\prime}=\frac{\delta \Gamma(q,-\omega_{q})}{\delta B}$ and 
$n_{b} (=\frac{1}{N}\sum_i \langle a^{\dagger}_i a_i\rangle$) is the density of magnons at the critical point. 
Let us define the following expressions
\begin{eqnarray}
&&\lambda \equiv \frac{A^{c}_{Q_{AF}}\Gamma^{c}_0}{8\pi^{2} c^{3}},\nonumber\\
&&\sigma  \equiv \Gamma^{\prime}n_{b}.
\label{e1040}
\end{eqnarray}
After some calculations we finally get the following relation
\begin{eqnarray}
 \Delta^{2}=\frac{(g\mu_{B}+\sigma)\delta B}{\lambda}\Big(1-\frac{ A^{c}_{Q_{AF}}\Gamma^{c}}{4\pi c^{2}}ln\frac{\delta B}{J}\Big).
\label{e1050}
\end{eqnarray}
The gap exponent $\phi$ is obtained upon replacing  $\Delta=|\delta B|^{\phi}$ in the 
above equation which finally is equal to
\begin{eqnarray}
 \phi=\frac{1}{2}-\frac{A^{c}_{Q_{AF}}\Gamma^{c}_0}{8\pi c^{2}}.
\label{e1060}
\end{eqnarray}
We have presented the numerical results of gap exponent in terms of anisotropic parameters 
in  section-\ref{summary}. 

\section{staggered magnetization for $B\lesssim B_c$}
At the quantum critical point the magnons condensate at $q=Q_{AF}$ which is the onset of the
long range antiferromagnetic (AF) order in the model. The system is represented by the 
AF ordered state as far as $B<B_c$. In the AF phase 
we have applied Hartree-Fock-Popov mean field approach \cite{shi,nikuni} by 
taking into account the condensation of magnons in the interacting Hamiltonian at wave vector $Q_{AF}$.
The effective interparticle interaction is defined by the following Hamiltonian
\begin{eqnarray}
\mathcal{H}_{eff}=\Gamma^{c}(Q_{AF})\sum_{k,k',q}a^{\dag}_{
k+q}a^{\dag}_{k'-q }a_{k'}a_{k} \;,
\label{e1070}
\end{eqnarray}
where, $\Gamma^{c}(Q_{AF})$ is the interaction parameter at the critical point ($B=B_{c}$). 
Below the critical point ($B< B_c$), the AF order parameter becomes nonzero and it can be interpreted by 
the nonzero mean field value of the creation operator of magnons at $Q_{AF}$.
Let us define $\langle a_{Q_{AF}}\rangle=N_c=\langle a^{\dagger}_{Q_{AF}}\rangle$
for the condensate phase where $N_c$ is the number of condensed magnons.
The staggered magnetization in the x-y plane
which represents the long range AF order is denoted by 
$m_{\perp}\equiv m_{x}+im_{y}=g\mu_{B}\sqrt{Nn_{c}}$
where $N$ is the total number of sites and $n_c$ is the condensed magnon density.
The effective Hamiltonian in Eq.(\ref{e1070}) can be written in the following form where
the contribution from the condensate phase has been denoted by $H_{\mathcal{U}}^{0}$,
\begin{eqnarray}
H_{\mathcal{U}}&=&H_{\mathcal{U}}^{0}+H_{\mathcal{U}}^{2}+H_{\mathcal{U}}^{3}+H_{\mathcal{U}}^{4},\nonumber\\
H_{\mathcal{U}}^{0}&=&\frac{\Gamma^{c}(Q_{AF})N_{c}^{2}}{2N},\nonumber\\
H_{\mathcal{U}}^{2}&=&\frac{\Gamma^{c}(Q_{AF})N_{c}}{N}\sum_{q}'\Big[\frac{1}{2}(a_{q}a_{-q}+a^{\dag}_{q}a^{\dag}_{-q})+2a^{\dag}_{q}a_{q}\Big],\nonumber\\
H_{\mathcal{U}}^{3}&=&\frac{\Gamma^{c}(Q_{AF})\sqrt{N_{c}}}{N}\sum_{k,q}'\Big(a^{\dag}_{k}a_{k+q}a_{-q}+h.c.\Big),\nonumber\\
H_{4}&=&\frac{\Gamma^{c}(Q_{AF})}{2N}\sum_{k,q,k^{'}}'a^{\dag}_{
k+q}a^{\dag}_{k'-q }a_{k'}a_{k}.
\label{e1080}
\end{eqnarray}
In the above equations, $\sum^{'}$ implies that the terms with creation and annihilation operators 
at the antiferromagnetic wave vector ($Q_{AF}$) are excluded. 
In a mean field approximation the contribution form $H_{\mathcal{U}}^{3}$ is zero since it
contains linear terms of boson operators.
Taking into account the hard core repulsion which avoids the pairing of magnons and considering
all other contractions the mean field representation of $H_{\mathcal{U}}^{4}$ is
\begin{eqnarray}
H_{\mathcal{U}}^{4}=2(1-n_{c})\Gamma^{c}(Q_{AF})\sum_{k}'a^{\dag}_{k}a_{k}.
\label{e1090}
\end{eqnarray}
After adding the non-interacting part, Eq.(\ref{e6}), to the mean field (MF) interacting one the 
Hamiltonian is given by the following equation plus a constant term which has been omitted here, 
\begin{eqnarray}
H^{MF}=\sum_{k}'(A_{k}+2\Gamma^{c}(Q_{AF}))a^{\dag}_{k}a_{k}+\sum_{k}'\frac{\nu J_{k}+\Gamma^{c}(Q_{AF})n_{c}}{2}(a_{k}a_{-k}+a^{\dag}_{k}a^{\dag}_{-k}).
\label{e1100}
\end{eqnarray}
The mean field Hamiltonian is diagonalized by the unitary Bogoliubov transformation,
\begin{eqnarray}
H^{MF}&=&\sum_{k}'\Omega_{k}(\phi^{\dag}_{k}\phi_{k}),\nonumber\\
\Omega_{k}&=&\sqrt{(A_{k}+2\Gamma^{c}(Q_{AF}))^{2}-(\nu J_{k}+\Gamma^{c}(Q_{AF})n_{c})^{2}},\nonumber\\
a_{k}&=&d_{k}\phi_{k}-f_{k}\phi^{\dag}_{-k},
\end{eqnarray}
where $\Omega_{k}$ gives the excitation spectrum of the new bosonic quasi partiles defined by
the creation operator $\phi^{\dag}_{k}$ and $d_{k}$, $f_{k}$ are Bogoliubov coefficients 
\begin{eqnarray}
d_{k}=\sqrt{\frac{A_{k}+2\Gamma^{c}(Q_{AF})}{2\Omega_{k}}+\frac{1}{2}},\;\;\;\ f_{k}=\sqrt{\frac{A_{k}+2\Gamma{c}(Q_{AF})}{2\Omega_{k}}-\frac{1}{2}}.
\end{eqnarray}
The condensation of magnons at the AF wave vector implies that excitation spectrum
must be gapless at $Q_{AF}$ which gives the following relation
\begin{eqnarray}
J_{\bf k=Q_{AF}} - J_0^z +g\mu _{B}B+2\Gamma^{c}(_{AF})=-\nu J_{k=Q_{AF}}-\Gamma^{c}(Q_{AF})n_{c}.
\label{e1150}
\end{eqnarray}
Therefore, the transverse order parameter for $B\leq B_{c}$  is given by
\begin{eqnarray}
m_{\perp}=g\mu_{B}\sqrt{n_{c}}=g\mu_{B}\sqrt{\frac{-J_{\bf k=Q_{AF}}+J_0^z -g\mu _{B}B-2\Gamma^{c}(Q_{AF})-\nu J_{Q_{AF}}}{\Gamma^{c}(Q_{AF})}}
\label{1151}
\end{eqnarray}
The critical field is touched where $n_{c}=0$ in  Eq.(\ref{e1150}) which leads 
to the following expression,
\begin{eqnarray}
B_{c}=\frac{J_{0}^{z}-J_{\bf Q_{AF}}-2\Gamma^{c}(Q_{AF})-\nu J_{\bf Q_{AF}}}{g\mu_{B}}.
\end{eqnarray}
Therefore the transverse staggered magnetization has the following expression in the
mean field approximation 
\begin{eqnarray}
m_{\perp}=g\mu_{B}\sqrt{n_{c}}=g\mu_{B}\sqrt{\frac{g\mu _{B}(B_{c}-B)}{\Gamma^{c}(Q_{AF})}}.
\label{tm}
\end{eqnarray}
The scaling behavior of transverse order parameter close to the critical field is characterized
by the exponent $\beta$ via $m_{\perp} \sim |B_c-B|^{\beta}$ which is $\beta=0.5$, in the mean field approximation.

\section{Summary, results and discussions \label{summary}}

In this article we have studied the effect of in-plane anisotropy on the quantum critical 
properties of the spin 1/2 Heisenberg model in the presence of a longitudinal field ($B$) on a
cubic lattice. The in-plane anisotropy breaks the U(1) symmetry of the model (around the direction
of the magnetic field) and changes the quantum critical properties of the model which is discussed
in this section. Moreover, we have analyzed the field-induced phase transition in this model 
in terms of Bose-Einstein condensation of magnons.

\begin{figure}
\vspace{1cm}
\includegraphics[width=10cm]{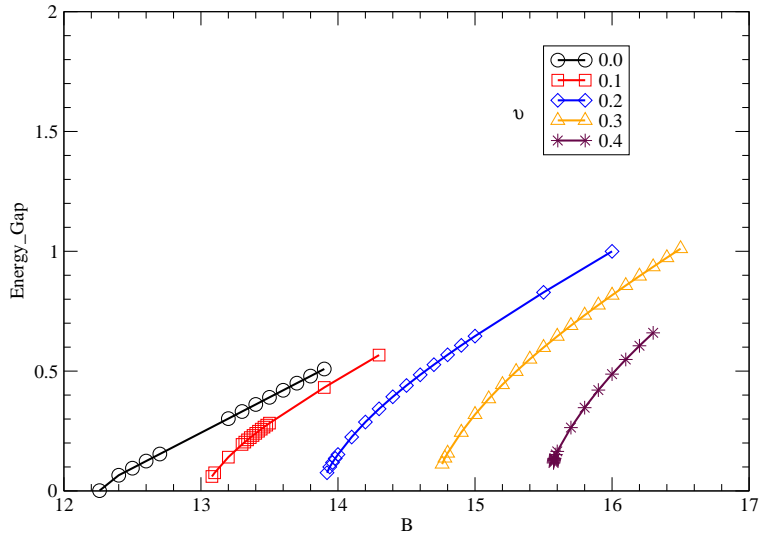}
\caption{\label{fig1} (Color online) The energy gap versus the magnetic field for various anisotropy parameters. The change 
of the critical magnetic field (where the gap vanishes) for various anisotropies is remarkable.}
\end{figure}

The original spin model has been represented by a bosonic model in the presence of hard core repulsion
to avoid double occupation of bosons at each lattice site which preserves the SU(2) algebra
of the spin model. 
In the limit of $B/J\longrightarrow\infty$, the ground state is a field induced ferromagnetic 
state and a finite energy gap exists to the lowest excited state which is called the magnon spectrum. 
The decrease of magnetic field lowers the excitation gap which eventually vanishes at
the critical magnetic field ($B_c$). This point corresponds to the
condensation of magnons which is the onset of long range antiferromagnetic order of the spin model.
We have implemented the Green's function approach to obtain the effect of interaction 
on the diagonal part of the bosonic Hamiltonian using Brueckner formalism close to the quantum 
critical point where the magnon density is small.
The magnon spectrum have been calculated from 
Eqs.(\ref{e330}, \ref{e340}, \ref{e211})
selfconsistently.
The procedure is started with an initial guess for $Z_{k}, \Sigma_{n}(k,0)$ and $\Sigma_{a}(k,0)$, 
then  using Eq.(\ref{e340}) we find the renormalized excitation energy and the 
renormalized Bogoliuobov coefficients. The procedure is repeated until convergence is reached. 
Using the final values for energy gap, renormalization constants and Bogoliubov coefficients, 
we have obtained the quantum critical point 
for different anisotropy parameters in Table.\ref{t1}. Our data shows that a small amount 
of anisotropy has a considerable change on the critical magnetic field. We have also plotted
the magnon gap versus the magnetic field in Fig.\ref{fig1} for different values of anisotropy.
It is obvious from Fig.\ref{fig1} that the magnon gap vanishes as the magnetic field approaches
the critical value for $B\gtrsim B_c$. 

Moreover, the scaling behavior of gap close to $B_c$ which is characterized by the
gap exponent  ($\phi$) defined
in Eq.(\ref{gapscaling}) depends
on the anisotropy parameter. We have presented the gap exponent ($\phi$) 
for different anisotropies in Table.\ref{t1}. The dependence of $\phi$ on 
$\nu$ shows that the in-plane anisotropy changes the universal behavior of the model.
A drastic change of $\phi$ from $0.40$ for $\nu=0$ to $0.2$ for $\nu=0.1$ manifests
the change of universality class due to explicit breaking of symmetry by anisotropy.
At $\nu=0$ the model has U(1) symmetry while for $\nu \neq 0$ the symmetry breaks to Z2.
Although the calculated gap exponent for different nonzero anisotropies 
may change slightly, 
the amount of changes are in the order of error bar which mainly comes from the
error bar inherited in the value of $B_c(\nu)$, nonzero values of $\nu$
do not present different universal behavior.

\begin{center}
\begin{table}[ht]
\caption{\label{t1} The critical magnetic field ($B_{c}$) and gap exponent ($\phi$) for different values 
of anisotropies. The error bar for all data is $\pm 0.05$. 
}\vspace{0.3cm}
\begin{ruledtabular}
\begin{tabular}{cccccccc}
$\nu$ &0.0 &0.1&0.2&0.3&0.4\\
\hline
$B_c$ (critical field)&12.26&13.03&13.88&14.72&15.53\\
$\phi$ (gap exponent)&0.40&0.2&0.2&0.2&0.2\\
\end{tabular}
\end{ruledtabular}
\end{table}
\end{center}

\begin{figure}
\vspace{1cm}
\includegraphics[width=10cm]{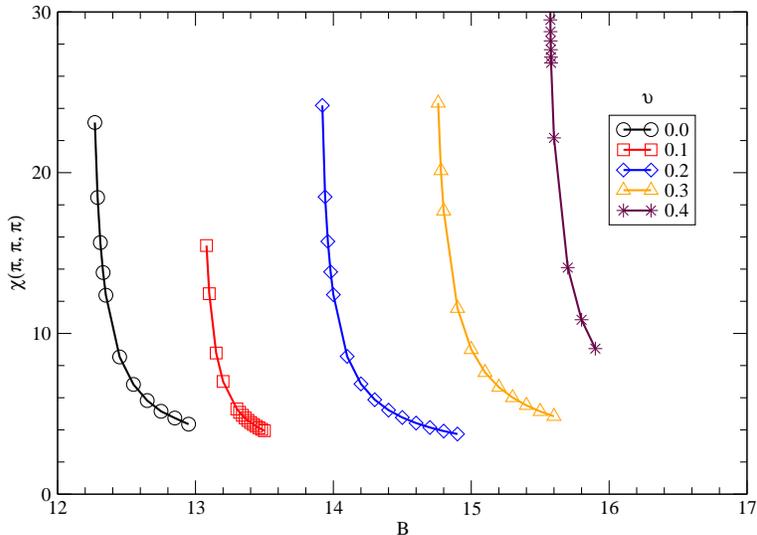}
\caption{\label{fig2} (Color online) The (x-component) transverse structure factor at the antiferromagnetic 
wave vector versus the magnetic field for different anisotropies. The divergence at the critical 
magnetic field justifies the onset of magnetic order.}
\end{figure}

The long range ordering can be deduced from the behavior of spin susceptibility. 
The magnetic ordering occurs for transverse spin components; therefore, static spin structure factor for tranverse component diverges at the antiferromagnetic wave vector in the quantum critical 
point.
The x-component static spin structure factor at momentum $q$ is defined by
\begin{eqnarray}
\chi^{xx}(q)=\langle s^{x}(q)s^{x}(-q)\rangle,
\label{211.7}
\end{eqnarray}
which is given by the following expression,
\begin{eqnarray}
\chi^{xx}(q)=\frac{\pi}{2}[2\sqrt{\frac{A_{q}^{2}}{4\omega_{q}^{2}}-1}+\frac{A_{q}}{\omega_{q}}].
\label{susceptibility}
\end{eqnarray}
Close to the quantum critical point where the magnon spectrum
vanishes the dominant contribution is obtained to be
\begin{equation}
\chi^{xx}(Q_{AF})\approx \pi A^{c}_{Q_{AF}} |B-B_{c}|^{-\phi},
\end{equation}
which shows that the divergens of magnetic susceptibility follows a scaling relation
with exponent, $\phi$.
The numerical results for $\chi^{xx}(Q_{AF})$ versus the magnetic field have been plotted
in Fig.\ref{fig2}. This plot confirms the presence of antiferromagnetic order at the
critical magnetic field. The divergence of the static structure factor happens at the
different critical fields for various anisotropies ($\nu$) which justifies the previous
results on energy gap. 

Both results on the energy gap and transverse structure factor confirms that a small amount of
in-plane anisotropy changes the critical magnetic field considerably. Moreover, the explicit
breaking of symmetry due to in-plane anisotropy shows up in the gap exponent. Our results
on the scaling behavior of magnetic order parameter ($\beta$) is limited to the mean field
approximation which does not show its dependence on the anisotropy. However, we expect that
the dependence of $\beta$ on anisotropy ($\nu$) should appear if the calculation goes
beyond the mean field approach. 
As far as the model has two control parameter $\nu$ and $B$, the universal
behavior should be fixed by two exponents $\phi$ and $\beta$. In other words, any other
exponent for a scaling behavior close to critical field at zero temperature can be expressed
in terms of the obtained exponents ($\phi, \beta$).

\section{acknowledgment}
We would like to express our deep gratitude to P. Thalmeier and V. Yushankhai 
who originally suggested
this problem and also for their valuable comments and fruitful discussions.
The authors would like to thank the hospitality of physics department of the institute for research
in fundamental sciences (IPM)  during part of this collaboration.
This work was supported in part by the Center of Excellence in
Complex Systems and Condensed Matter (www.cscm.ir).


\end{document}